\documentclass[12pt]{article}

\usepackage{amsfonts}
\usepackage{amssymb}
\usepackage{amsmath}

\usepackage{amsthm}

\usepackage{authblk}

\addcontentsline{toc}{section}{Appendices}

\newcommand{\bm}[1]{\mbox{\boldmath $#1$}}

\def\defi{:=}

\def \a {\alpha}
\def \b {\beta}
\def \m {\mu}
\def \n {\nu}
\def \r {\rho}
\def \g {\gamma}
\def \l {\lambda}

\def \d {\delta}

\def \N {\nabla}
\def \Nb {\overline{\nabla}}

\def\deltaS{\delta^\Sigma}

\def\RS{R^\Sigma}

\def\S{\Sigma}

\def\otheta{\underline{\theta}}
\def\1{\underline{1}}

\def\be{\begin{equation}}
\def\ee{\end{equation}}
\def\bea{\begin{eqnarray}}
\def\eea{\end{eqnarray}}
\def\bean{\begin{eqnarray*}}
\def\eean{\end{eqnarray*}}

\def\={\stackrel{\Sigma}{=}}

\newlength{\cellwidth}
\begin{document}

\title{Junction Conditions for General Gravitational Theories}
\author{Jos\'e M. M. Senovilla\\
Departamento de F\'isica, Universidad del Pa\'is Vasco UPV/EHU, Apartado 644, 48080 Bilbao, Spain\\
and\\
EHU Quantum Center, Universidad del Pa\'{\i}s Vasco UPV/EHU, Bilbao, Spain.}

\maketitle

\vspace{-0.2em}

\begin{abstract}
The junction conditions for general theories of gravity based on actions that depend on arbitrary functions of the curvature scalar invariants (including differential invariants) are obtained using the distributional formalism. In case of the existence of thin shells, a general expression for the shell energy-momentum tensor is presented. Generalized Israel equations are also obtained. The conditions for a proper matching, without shells, are derived. The main results are: (i) shells arise if the $m$th-covariant derivative of the Riemann tensor is continuous at the matching hypersurface, where $m$ is the maximum order of differentiation appearing in the Lagrangian density; (ii) a proper junction without thin shells requires further that the $(m+1)$-th derivative be also continuous, (iii) theories with $m=0$ that are quadratic in the scalar curvature invariants are special and unique for they allow for discontinuities of the Riemann tensor resulting in the existence of thin shells and {\em gravitational double layers} and (iv) General Relativity and $F(R)$ theories are extraordinary theories that admit shells of curvature (i.e. impulsive gravitational waves) because other theories require the absence of jumps of the second fundamental form across the matching hypersurface. For proper junctions, the continuity across the matching hypersurface of the normal components of the energy-momentum tensor is proven to be a {\em universal} property, independently of the field equations, thereby providing important necessary conditions for any matching in any gravitational theory. All results are derived for a minimal coupling with the matter, but the strategy would be analogous for more general couplings. 
\end{abstract}

\section{Introduction}
Gravitational surface layers or thin shells are idealized objects that describe localized concentration of energy on a given hypersurface. They represent domain walls, brane-worlds \cite{MK}, impulsive waves, thin layers of matter or of gravitational fields, concentrated lightlike signals, propagation of null matter, etcetera. In General Relativity, pioneering work by Lanczos \cite{La,La1}  was followed by the well-known and fundamental {\em Israel equations} \cite{I} for the case of non-null shells. This was later extended to the study of null shells in \cite{CD,BI,BH,BH1} and to general shells in \cite{MS,S4} with applications in \cite{MSV,MSV1,MSV2}.

The purpose of this paper is to derive the general junction conditions for arbitrary gravitational theories. To that end, in this paper I restrict attention to the case of timelike matching hypersurfaces, the case of primary interest. Today there is a plethora of alternative theories, based on all kind of gravitational actions. However, there seem to be difficulties in identifying the junction conditions in some complicated situations \cite{BCHM,BCHM1,BCHMV}, and sometimes the junction conditions used may not be the correct ones. In this work, the general junction conditions, for any theory based on actions that depend arbitrarily on the metric, the curvature tensor and (possibly) its covariant derivatives are found.\footnote{The analysis is restricted to metric theories of gravity. Actions that depend on torsion and/or non-metricity are not considered in this paper.}  The guiding principle to derive them is simply that the field equations must make sense in the distributional sense, thereby allowing for the existence of Dirac-delta distributions with support on the matching hypersurface. Hence, I use the formalism of tensor distributions in Lorentzian geometry, for details consult \cite{I,L,L1,T,GT,P,Poi,MS,RSV}.

A general expression for the energy-momentum tensor on the shell is presented (Eqs.\eqref{tau1} and  \eqref{tau2}). This tensor is proven to be tangent to the hypersurface and satisfy generalized Israel equations relating its divergence (within the shell) with the discontinuities of the bulk energy-momentum tensor: Eqs.\eqref{genIsrael}. This implies, for a proper matching, the traditional continuity of the normal components of the energy-momentum tensor, Eq.\eqref{nT}, that can be used as necessary matching properties in arbitrary junctions.

Along the way, it turns out that General Relativity, $F(R)$ theories, and quadratic theories are very special. The latter are the only ones allowing for gravitational double layers \cite{Senovilla13,S2,S3,RSV,EGS}, while the two previous permit discontinuities of the second fundamental form on the matching hypersurface. 

There have been some previous contributions relevant to this work, as for instance \cite{RM} where the junction conditions in quadratic theories were derived using the boundary term approach. The results agree with \cite{RSV} except for the double layers, that do not arise using the boundary approach. Therefore, doubts about the generality of the boundary approach arise. Notice that this was the approach used in \cite{BCHM,BCHM1,BCHM2,BCHMV}, and it is not clear to me that their results comply with the general junction conditions presented herein. In \cite{Rosa} some particular theories depending on a selection of invariants was analyzed and the junction conditions found using the distributional approach. In \cite{Rac} a very general method to derive junction conditions of general fields other than gravity was presented.

The paper is structured as follows. In section \ref{sec:basic} some basic features of the matching problem are collected, to fix the notation and the fundamental objects to be used. For readers not familiar with the distributional approach, this section should be combined with the Appendix, in which I collect all relevant formulas to be used in the main text with some explanations. More details can be found in \cite{MS,RSV}. In section \ref{sec:F} the general theory based on arbitrary actions that depend on the curvature scalar invariants is studied and the main results derived. Particular subsections are devoted to the specific cases of General Relativity, $F(R)$ and quadratic Lagrangian densities. The case with Lagrangian densities that depend on the metric, the curvature {\it and} its covariant derivatives of any order is then considered in section \ref{sec:hatF}. There are some claims that these theories are ill-defined, from a (quantum) causality viewpoint \cite{ESR}, but nevertheless they can be considered in a general setting \cite{IW} and I have decided to include their junction conditions for completeness. Finally, a brief discussion is placed before the Appendix.

\section{The basic gluing}\label{sec:basic}
Let $(M,g)$ denote a connected, oriented and time-oriented $(n+1)$-dimensional Lorentzian manifold with metric $g$ of signature $(-,+,\dots,+)$. Consider the case where $(M,g)$ has two different regions, possibly with different matter and/or different gravitational fields, separated by a timelike hypersurface $\S\subset M$. Lowercase Greek letters $\alpha,\beta,\dots $ are spacetime indices and run from $0$ to $n$. Small Latin indices $a,b,\dots$ are hypersurface indices and take values from $1$ to $n$. Both sides of $\S$ are denoted by $M^+$ and $M^-$, with corresponding smooth metrics $g_{\m\n}^+$ and $g_{\m\n}^-$ that have well defined, definite limits, when approaching $\S$. It is known  \cite{MS,Poi} that one can construct a local coordinate system such that the global metric is continuous if, and only if, the first fundamentals forms inherited by $\S$ from $M^+$ and $M^-$ agree. This is a basic prerequisite that is indispensable to have a well-defined spacetime structure.  

Let $\{\xi^a\}$ be a set of local coordinates on $\Sigma$ ($a,b,\dots =1,\dots , n$). Then, there are two parametric representations 
$$
x_{\pm}^\mu =x_{\pm}^\mu (\xi^a)
$$ 
of $\Sigma$, one for each embedding into each of $M^\pm$.  The agreement of the inherited metrics amount to the following equalities on $\Sigma$
\be
h^+_{ab} = h_{ab}^- ,\hspace{1cm}
h^\pm_{ab} \equiv g_{\mu\nu}^\pm (x(\xi))\frac{\partial x_\pm^\mu}{\partial \xi^a}\frac{\partial x_\pm^\nu}{\partial \xi^b} .
\label{h=h}
\ee
The unique continuous metric that can then be defined on the entire manifold and coincides with
$g^{\pm}$ in the respective $M^{\pm}$ is denoted simply by $g_{\mu\nu}$.

Let $n^\pm_\mu$ be the unit normals to $\Sigma$ as seen from $M^\pm$ respectively. They are fixed up to a sign by the conditions
$$
n^\pm_\mu \frac{\partial x_\pm^\mu}{\partial \xi^a}=0, \quad n^\pm_\mu n^{\pm\mu}=1
$$
and one must choose (say) $n^-_\mu$ pointing outwards from $M^-$ and $n^+_\mu$ pointing towards $M^+$. Hence, the two bases on the tangent spaces at any point of $\S$
$$
\{n^{+\mu},\frac{\partial x_+^\mu}{\partial \xi^a}\} \quad \quad \leftrightarrow \quad \quad\{n^{-\mu},\frac{\partial x_-^\mu}{\partial \xi^a}\}
$$ 
agree and are then identified, the $\pm$ are dropped (even though, in explicit calculations, one can still use both versions using the two coordinate systems on each side). Then, a unique unit normal $n_\mu$, that points from $M^-$ towards $M^+$, is defined so that
$$
n_\mu e^\mu_a =0
$$
where $\{\vec{e}_a\}$ are the vector fields tangent to $\S$ defined by the above embeddings
$$
\vec{e}_a \defi \left.\frac{\partial x_+^\mu}{\partial \xi^a}\frac{\partial}{\partial x^\mu_+}\right|_\S =\left.\frac{\partial x_-^\mu}{\partial \xi^a}\frac{\partial}{\partial x^\mu_-}\right|_\S .
$$
Note that $\{\vec{e}_a\}$ are defined only on $\S$. The basis dual to $\{n^\mu,e^\mu_a\}$ is denoted by
$$
\{n_\mu, \omega^a_\mu\}
$$
where the one-forms $\bm{\omega}^a$ are characterized by
$$
n^\mu \omega^a_\mu =0, \hspace{1cm} e^\mu_b \omega^a_\mu =\delta^a_b .
$$

From this point on, the connection and curvature of $(M,g)$ can be computed in a distributional sense. The main necessary formulas are collected here in the Appendix, and I refer to \cite{MS,RSV} for the explicit calculations and detailed explanations.

\section{Gravitational theories based on scalar invariants of the curvature: junction conditions}\label{sec:F}
Let us consider a general theory of gravity in $n+1$ dimensions described by the Lagrangian density
\begin{equation}
\mathcal{L} = \frac{1}{2\kappa} F(\mbox{Riemann scalar invariants})+\mathcal{L}_{matter},\label{lag}
\end{equation}
where $\kappa =8\pi G/c^4$ is the gravitational coupling constant, $F$ is any function that can depend on all possible curvature scalar invariants, of any degree, of the Riemann tensor, and $\mathcal{L}_{matter}$ is the Lagrangian density describing the matter fields. The function $F$ can be a polynomial of the curvature scalar invariants, but can also be expandable as an infinite power series of the different scalar invariants of any degree. Of course, the above general case includes General Relativity (GR), $F(R)$ theories, quadratic or cubic  theories, Lanczos-Lovelock,  as well as the so-called generalized quasi topological gravities \cite{BCH,BCH1,BCHLM,BCMM,MM}, among many others.

The field equations derived from this Lagrangian read (see e.g. \cite{Pad,Pad1,BCMM,BCH,BCHLM} and references therein)
\be\label{FE}
P_\a{}^{\rho\g\d} R_{\beta\rho\g\d}-\frac{1}{2} F g_{\a\b} -2\nabla_\rho\nabla_\g P_\a{}^{\rho\g}{}_\beta=\kappa T_{\alpha\beta}
\ee
where
\be
P^{\a\b\g\d} := \frac{\partial F}{\partial R_{\a\b\g\d}} \, .
\ee

The goal in this paper is to obtain the correct junction conditions when there are regions of the spacetime with different matter content separated by a timelike hypersurface where $T_{\alpha\beta}$ can be discontinuous or even have a thin shell of matter. The guiding principle to obtain such junction conditions is simply to find the requirements that render the field equations \eqref{FE} mathematically well-defined: even allowing for distributional tensors, they must make sense. The discussion can be split into the general case, theories involving only up to quadratic invariants, and GR.

\subsection{The generic case}
The generic case is defined as one in which quadratic or higher-degree terms appear in the expansion of the function $F$. 
\begin{itemize}
\item The field equations will have quadratic or higher degree summands coming from $F g_{\a\b}$ and from $P_\a{}^{\rho\g\d} R_{\beta\rho\g\d}$. These do not make mathematical sense within the distributional theory if they contain multiplication of singular distributions such as products of the $\deltaS$'s. Recalling the general formulas \eqref{Riedist} and \eqref{HRie} this leads to the indispensable requirement 
\be\label{Kdisc=0}
\left[K_{\mu\nu}\right] =0 .
\ee
In short, the second fundamental form must be the same as inherited from both sides of the hypersurface. Then the Riemann tensor 
$$
\underline{R}^\alpha{}_{\beta\mu\nu}=R^{+\alpha}{}_{\beta\mu\nu}\otheta + R^{-\alpha}{}_{\beta\mu\nu}(\1-\underline{\theta})
$$
is a distribution associated to a locally integrable tensor field, with at most a finite jump across $\S$, and all products of the curvature tensor with itself make good sense.
\item Assuming the necessary requirement \eqref{Kdisc=0}, the remaining term in \eqref{FE} will contain summands of type (Rie represents the Riemann tensor schematically)
\begin{center}
Rie$^{k}\,  \nabla$Rie $\nabla$Rie , \hspace{1cm}  Rie$^k\,  \nabla\nabla$Rie
\end{center}
and those on the left present again the same problem of multiplication of singular distributions of type $\deltaS \deltaS$, according to \eqref{nablaRie1}, that cannot be given a good mathematical interpretation. The only exception is the case where those terms on the left do not arise, and this can only happen if the theory is purely polynomially quadratic. This case is briefly considered in the next subsection. For the generic case, however, one needs the vanishing of the discontinuity of the Riemann tensor 
\be\label{Riedisc=0}
\left[R_{\a\b\g\d}\right] =0
\ee
and thus the Riemann tensor field must be continuous across $\S$. Observe that this requires $B_{\mu\nu}=0$, due to \eqref{eq:[Rie]} and thus from (\ref{nablanablaRie}--\ref{eq:rel_tB}) we have 
\bea
\nabla_\tau \nabla_\l \underline{R}_{\a\beta\mu\nu}=\nabla_\tau\nabla_\l R^{+}_{\a\b\m\n} \otheta +\nabla_\tau\nabla_\l R^{-}_{\a\b\m\n} (\1-\otheta) \nonumber \\
+ (n_\a n_\m \r_{\b\n} -n_\a n_\n \r_{\b\m} - n_\b n_\m \r_{\a\n}+ n_\b n_\n \r_{\a\m}) n_\l n_\tau \deltaS \label{DDRiem}
\eea
where
$$
[\nabla_{\rho}R_{\alpha\beta\mu\nu}]=n_\rho (n_\a n_\m \r_{\b\n} -n_\a n_\n \r_{\b\m} - n_\b n_\m \r_{\a\n}+ n_\b n_\n \r_{\a\m}), \hspace{7mm} \r_{\b\n}= n^\rho n^\a n^\mu [\nabla_{\rho}R_{\alpha\beta\mu\nu}].
$$
\end{itemize}
The cases with $\rho_{\a\b}\neq 0$ admit thin shells of matter, because the field equations \eqref{FE} will require an energy-momentum tensor of type
\be\label{e-m}
\underline{T}_{\a\b} = T_{\a\b}^+ \otheta +T^-_{\a\b} (\1-\otheta) + \tau_{\a\b} \deltaS
\ee
to match the terms involving $\rho_{\a\b}$ multiplying $\deltaS$ coming from the $\nabla^\rho\nabla^\g P_{\a\rho\g\beta}$ part of the field equations. 

The particular expression for $\tau_{\a\b}$ depends on the particular theory, that is, on the particular expression of the Lagrangian density function  $F$. It basically collects all the terms that appear in $\nabla^\rho\nabla^\g P_{\a\rho\g\beta}$ containing a second covariant derivative of the curvature. 
From the definition of $P_{\a\rho\g\beta}$ one readily deduces that $\nabla_\rho\nabla_\g P_\a{}^{\rho\g}{}_\beta$ has the following structure
\be\label{defS}
\nabla_\rho\nabla_\g P_\a{}^{\rho\g}{}_\beta = S_{\alpha\beta}{}^{\tau\l\g\delta\m\n}\nabla_\tau \nabla_\l R_{\g\delta\m\n}+\mbox{terms}(\nabla R, R)
\ee
where $S_{\alpha\beta}{}^{\tau\l\g\delta\m\n}$ is a tensor field that depends on the Riemann tensor and the metric but {\em not} on its derivatives and the rest of the terms do not involve any 2nd-order derivatives of the Riemann tensor field. Hence, on using \eqref{DDRiem} one obtains the formula for the energy-momentum tensor on the thin shell:
 \be\label{tau}
 \tau_{\a\b}:=-8 \, \, n_\tau n_\l n_\g n_\mu\,  \rho_{\delta\n}\, \,  S_{\alpha\beta}{}^{\tau\l\g\delta\m\n}|_\S
 \ee
 or more informatively
 \be\label{tau1}
 \tau_{\a\b}:=-2 \, \, n_\tau n_\l n_\g n_\mu\,  n^\rho n^\sigma n^\epsilon [\nabla_{\rho}R_{\sigma\delta\epsilon\nu}]\, \,  S_{\alpha\beta}{}^{\tau\l\g\delta\m\n}|_\S \, .
 \ee

For each explicit theory with $F$ known, $\tau_{\a\b}$ can be readily computed by using the definition \eqref{defS} of $S_{\alpha\beta}{}^{\tau\l\g\delta\m\n}$. Furthermore, despite not having its explicit expression for general $F$, one can prove that $\tau_{\a\b}$ is necessarily tangent to $\S$ and that a generalized form of the Israel equations holds. To that end, we rely on the covariant conservation of the righthand side of the field equations \eqref{FE} \cite{IW,Peng,PL}, and using the conservation of $T^\pm_{\a\b}$ on $M^\pm$ respectively and the relations \eqref{nablaT1} and \eqref{nablaTdelta} one gets
$$
0= \nabla^\a \underline{T}_{\a\b} = n^\a \left[T_{\a\b}\right] \deltaS + \left(h^{\a\sigma}\nabla_\sigma \tau_{\a\b} -K^\rho{}_\rho n^\a \tau_{\a\b} \right)\deltaS +\nabla_\rho(\tau_{\a\b} n^\a n^\rho \deltaS)
$$ and the last term cannot cancel any other part, so that necessarily 
\be\label{tautangent}
n^\a \tau_{\a\b} =0 .
\ee
Therefore, one can actually re-write \eqref{tau} as
\be\label{tau2}
 \tau_{\a\b}:=-8 \, \, n_\tau n_\l n_\g n_\mu\,  \rho_{\delta\n}\, \,  h^\rho_\a h^\epsilon_\b \, \, S_{\rho\epsilon}{}^{\tau\l\g\delta\m\n}|_\S
 \ee

Taking this tangential property into account together with
$$
h^\rho{}_\l \nabla_\r \tau_{\a\b} = \overline\nabla_\l \tau_{\a\b} -2K^\r{}_\l \tau_{\r(\a} n_{\b)}
$$
where $\overline\nabla$ is the covariant derivative within $(\S,h)$, the conservation equation finally implies
$$
0= \left(n^\a \left[T_{\a\b}\right]+\overline\nabla^\a \tau_{\a\b} - K^{\rho\sigma}\tau_{\r\sigma} n_\b \right)\deltaS 
$$
which, splitting into tangent and normal parts to $\S$, provide
\be\label{genIsrael}
n^\a h^\r{}_\b \left[T_{\a\r}\right]+\overline\nabla^\a \tau_{\a\b} =0, \hspace{1cm}  n^\a n^\b \left[T_{\a\b}\right] - K^{\a\b}\tau_{\a\b} =0.
\ee
These are the generalized Israel equations, as they keep exactly the same structure as the original ones \cite{I} except for the important point that now the explicit expression of the shell energy momentum tensor $\tau_{\a\b}$ depends crucially on the form of the Lagrangian function $F$. This explicit expression depends essentially on the discontinuities of normal derivatives of the Riemann tensor field, encoded into $\r_{\b\n}= n^\rho n^\a n^\mu [\nabla_{\rho}R_{\alpha\beta\mu\nu}]$.

The case of a {\em proper matching} without thin shells, just allowing for finite jumps of the energy-momentum tensor, require the extra condition $\tau_{\a\b}=0$. Generically this can only happen when $\rho_{\a\b}=0$, so that one also needs to require
\be\label{nablaRiedisc=0}
[\nabla_{\rho}R_{\alpha\beta\mu\nu}]=0
\ee
together with \eqref{Kdisc=0} and \eqref{Riedisc=0}. One must keep in mind, however, that there will always be extremely special theories (that is, very particular expressions of the Lagrangian function $F$) in which the vanishing of $\tau_{\a\b}$ does not require the full condition \eqref{nablaRiedisc=0} but just part of it. Known examples are Gauss-Bonnet, $F(R)$ or some very special quadratic theories, see e.g. \cite{RSV}. Independently of that, for a proper matching the continuity equations
\be\label{nT}
n^\a [T_{\a\b}] =0
\ee
always hold. In short, this proves the universality of the continuity of the $T_{\mu\nu}$-normal components across $\S$, which can thus be used as necessary matching conditions in arbitrary matchings. In particular, observe that matching with vacuum will require the vanishing of the normal pressure, and this will generically define the matching hypersurface in that situation.

\subsection{Purely quadratic cases}
As mentioned in the previous subsection, an exception arises if the theory is purely polynomially quadratic, so that the function $F$ takes the generic form
$$
F=-2\Lambda + R +a_1 R^2 +a_2 R_{\m\n} R^{\mu\nu} +a_3 R_{\a\b\m\n} R^{\a\b\m\n}
$$
for generic constants $\Lambda, a_1, a_2, a_3$ with $a_2^2+a_3^2\neq 0$. The same reasoning as before leads to \eqref{Kdisc=0}, but now $P^\a{}_{\b\g\d}$ is {\em linear} in the Riemann tensor and therefore the terms coming from the $\nabla^\rho\nabla^\g P_{\a\rho\g\beta}$ part of the field equations do not have ill-defined products and thus do not require \eqref{Riedisc=0} to hold. The derivative of the Riemann tensor distribution can take the form \eqref{nablaRie1}, and its second covariant derivative is expression \eqref{nablanablaRie}. The last summand in the latter expression implies the existence of {\em gravitational double layers} \cite{Senovilla13,S2,S3,RSV}, because the energy-momentum tensor distribution must necessarily take the form
$$
\underline{T}_{\a\b} =T_{\a\b}^+ \otheta +T^-_{\a\b} (\1-\otheta) + (\tau_{\a\b} +\tau_\a n_\b +\tau_\b n_\a +\tau n_\a n_\b) \deltaS +\underline{t}_{\a\b}
$$
where $\underline{t}_{\a\b}$ is the singular part of $\underline{T}_{\a\b}$ not proportional to $\deltaS$ and gives the double-layer part of the energy-momentum. Here, the tensor field multiplying $\deltaS$ has been orthogonally split into its part $\tau_{\a\b}$  tangent to $\S$, its part half tangent $\tau_\a$ and its fully normal part $\tau$:
$$
\tau_{\a\b}=\tau_{\b\a}, \hspace{4mm} n^\a \tau_{\a\b} =0, \hspace{4mm} n^\a \tau_\a =0.
$$
Thus, there arise new terms in the part proportional to $\deltaS$ that are not tangent to $\S$, described by the {\em external flux momentum} $\tau_\a$, and the {\em external pressure/tension} $\tau$. 

These purely quadratic theories where thoroughly analyzed in \cite{RSV}, where explicit expressions for all these quantities were provided and a proof of the full energy-momentum conservation given. In particular, the Israel equations must be modified to take the double layer into account, and the new relations were found to be
\begin{align}
n^\alpha h^\r_\b [T_{\a\r}]+ \Nb^\a \tau_{\a\b} 
&= -\mu_{\a\r} \overline{\N}_\b K^{\a\r} + \overline{\N}_\r (\m^{\a\r}K_{\a\b}) - \overline{\N}_\r(\m_{\a\b}K^{\a\r}) ,
 \label{eq:1} \\
n^\a n^\b [T_{\a\b}] - \tau_{\a\b}K^{\a\b} &=  
\Nb^\a\Nb^\b \m_{\a\b} + \mu^{\m\n}\left (n^\a n^\g \RS_{\a \m\g\n} + K_\m^\r K_{\n\r} \right ), \label{eq:2}\\
\tau_\a +\overline\nabla^\r\mu_{\r\a} &=0, \\
\tau -K^{\a\b}\mu_{\a\b} &=0
\end{align}
where $\mu_{\a\b}$ is a symmetric tensor field defined on $\S$ and tangent to $\S$ that measures the double-layer strength
$$
\kappa \mu_{\a\b}\defi (2a_1 +a_2+2a_3) [R] h_{\a\b} +(4a_3+a_2) [G_{\a\b}] .
$$
When the double-layer strength $\mu_{\a\b}$ is set to zero, one falls back into the generic situation of the previous subsection (in this case with known explicit formulas for $\tau_{\a\b}$) and, in particular, a proper matching requires again \eqref{Kdisc=0}, \eqref{Riedisc=0} and \eqref{nablaRiedisc=0} ---unless very special relations between the constants $a_1,a_2,a_3$ exist.

For details I refer to \cite{RSV}.

\subsection{GR}
General Relativity is an exceptional, and thereby different and unique, case because it is the only theory with a Lagrangian function $F$ linear in the curvature. Thus, the field equations are also linear in the curvature and there is no need even to assume \eqref{Kdisc=0}. Then, the standard expression for the energy-momentum on the shell follows from \eqref{HG} and the traditional Israel conditions can be deduced immediately from \eqref{1}-\eqref{2}. 

The proper matching in GR only requires \eqref{Kdisc=0}, as is well known.

\subsection{$F(R)$-theories as peculiar cases}
The particular cases where the Lagrangian density depends on the scalar curvature $R$ exclusively, known as $F(R)$ theories, are of a peculiar kind. This happens because the field equations \eqref{FE} read \cite{Senovilla13}
\be
F'(R)R_{\mu\nu}-\frac{1}{2}F(R)g_{\mu\nu}-\nabla_{\mu}\nabla_{\nu}F'(R)+g_{\mu\nu}\nabla_{\rho}\nabla^{\rho} F'(R) =\kappa T_{\mu\nu} \label{fe}
\ee
where primes denote derivatives with respect to $R$. These equations can be seen to become
\bean
F'(R)R_{\mu\nu}-\frac{1}{2}F(R)g_{\mu\nu}-F''(R)\left( \nabla_\mu \nabla_\nu R -g_{\mu\nu} \nabla_\rho\nabla^\rho R\right)\\
-F'''(R)\left(\nabla_\mu R \nabla_\nu R -g_{\mu\nu} \nabla_\rho R \nabla^\rho R \right)=\kappa T_{\mu\nu} 
\eean
where one notices that, apart from the typical term linear in the Ricci tensor, only products and derivative of the scalar curvature arise. Therefore, it is enough to have a well defined scalar curvature as an square integrable function and this only requires, according to \eqref{scalardist} and \eqref{Hscalar}, that the trace of the second fundamental form has no jump:
\be\label{traceK}
[K^\r{}_\r]=0 .
\ee
Then, explicit formulas for the shell energy-momentum tensor and, in the purely quadratic case, for the double-layer strength were given in \cite{Senovilla13}. {\em Observe that $F(R)$ theories, including GR, are the only cases that contain shells of curvature, or in simpler words,  impulsive gravitational waves}.

In these exceptional $F(R)$-theories, the proper matching without shells requires \eqref{Kdisc=0} together with
$$
[R]=0, \hspace{1cm} n^\rho[\nabla_\rho R]=0.
$$
For details I refer to \cite{Senovilla13,S2,S3}.

\section{General theory equations: junction conditions}\label{sec:hatF}
I turn now to the more general case of Lagrangian densities that are built out of scalar invariants formed with the metric tensor, the Riemann curvature {\em and} the covariant derivatives of the Riemann tensor. Again, both polynomial and infinite power series constructed on these scalar invariants are permitted:
\begin{equation}
\mathcal{L} = \frac{1}{2\kappa} \hat{F}(\mbox{scalar invariants from Riemann and its derivatives})+\mathcal{L}_{matter}.\label{lag1}
\end{equation}
Let $m$ denote the maximum order of the covariant derivative involved the previous expression, so that $\nabla_{\a_1}\dots \nabla_{\a_m}R_{\a\b\m\n}$ does appear in $\hat F$ ---and possibly many other covariant derivatives of order less than $m$.

The field equations can be written in the following schematic form \cite{Peng,PL,IW}:
\be\label{FE1}
\hat{P}_\a{}^{\rho\g\d} R_{\beta\rho\g\d}-\frac{1}{2} \hat{F} g_{\a\b} -2\nabla_\rho\nabla_\g \hat{P}_\a{}^{\rho\g}{}_\beta+E_{\a\b} =\kappa T_{\alpha\beta}
\ee
where now
$$
\hat{P}^{\a\b\m\nu} := \sum_{i=0}^m (-1)^i \, \nabla_{\a_1} \dots \nabla_{\a_i} \left(\frac{\partial \hat F}{\partial \nabla_{a_1} \dots \nabla_{\a_i} R_{\a\b\m\n}}\right)
$$
and $E_{\a\b}$ is a complicated expression involving many products of covariant derivatives of the Riemann tensor field of several orders; for explicit expressions  consult \cite{Peng,PL}. The important point here is that $E_{\a\b}$ contains such covariant derivatives {\em up to order} $m$, and not higher. Therefore, the lefthand side of the field equations \eqref{FE1} will contain many products of covariant derivatives of the curvature {\em up to order} $m+2$, and the summands containing the $(m+2)$-th derivative $\nabla_{\a_1}\dots \nabla_{\a_{m+2}} R_{\a\b\m\n}$ come exclusively from the term $\nabla_\rho\nabla_\g \hat{P}_\a{}^{\rho\g}{}_\beta$ in \eqref{FE1}.

Based on these facts, and on the discussion in the previous section, one knows that there will always be ill-defined terms in the field equations (coming already from $\hat F$ itself in the previous equation) unless the Riemann tensor has no singular part. Therefore, \eqref{Kdisc=0} is again absolutely necessary. But not only that, all products of covariant derivatives of the curvature tensor must not contain singular $\deltaS$-parts, and thus in addition to \eqref{Kdisc=0} one needs to require
\be\label{manydisc=0}
[\nabla_{\a_1}\dots \nabla_{\a_i} R_{\a\b\m\n}] =0 \hspace{1cm} \mbox{for all} \hspace{1cm} i\in\{0,1,\dots,m\} .
\ee
The Riemann tensor field must be $m$-times differentiable. The first possible discontinuity arises at the $m+1$-order, so that
\bea
\nabla_{\a_1}\dots \nabla_{\a_{m+1}} \underline{R}_{\a\b\m\n} = \nabla_{\a_1}\dots \nabla_{\a_{m+1}} R^+_{\a\b\m\n}\,  \otheta +\nabla_{\a_1}\dots \nabla_{\a_{m+1}} R^-_{\a\b\m\n}\, (\1-\otheta) \nonumber\\
\nabla_{\a_1}\dots \nabla_{\a_{m+2}} \underline{R}_{\a\b\m\n} = \nabla_{\a_1}\dots \nabla_{\a_{m+2}} R^+_{\a\b\m\n}\,  \otheta \hspace{4cm}\nonumber\\
+\nabla_{\a_1}\dots \nabla_{\a_{m+2}} R^-_{\a\b\m\n}\, (\1-\otheta) +\left[\nabla_{\a_2}\dots \nabla_{\a_{m+2}} R_{\a\b\m\n}\right] n_{\a_{1}} \deltaS \, .\label{cosa}
\eea

This last summand, proportional to $\deltaS$, gives rise to a thin shell of matter in order to fulfill the field equations \eqref{FE1}, and thus we have again the expression \eqref{e-m} for the energy-momentum tensor distribution. As in the previous section, the specific formula for its singular part $\tau_{\a\b}$ depends on the particular structure of the function  $\hat F$. In the present situation, it collects all terms arising in $\nabla^\rho\nabla^\g \hat{P}_{\a\rho\g\beta}$ containing a $(m+2)$-th covariant derivative of $R_{\a\b\m\n}$. 
From the definition of $\hat P_{\a\rho\g\beta}$ one readily deduces that $\nabla_\rho\nabla_\g P_\a{}^{\rho\g}{}_\beta$ has the following structure
\be\label{defS1}
\nabla_\rho\nabla_\g P_\a{}^{\rho\g}{}_\beta = \hat{S}_{\alpha\beta}{}^{\tau\l\r_1\dots\r_m\g\delta\m\n}\nabla_\tau \nabla_\l \nabla_{\r_1} \dots\nabla_{\r_m}R_{\g\delta\m\n}+\mbox{terms}(\nabla^{m+1} R, \nabla^mR,\dots,\nabla R, R)
\ee
where $\hat{S}_{\alpha\beta}{}^{\tau\l\r_1\dots\r_m\g\delta\m\n}$ is a tensor field that depends on the metric, the Riemann tensor and its covariant derivatives up to order $m+1$,  but {\em not} on its $(m+2)$-th covariant derivative, and the rest of the terms contain derivatives of the Riemann tensor field up to order $m+1$ at most. Hence, on using \eqref{cosa} one obtains the formula for the energy-momentum tensor on the thin shell:
 \be\label{tau2}
 \tau_{\a\b}:=-2 \, \, n_\tau n_\l n_\g n_\mu\, n^\r n^\sigma n^\epsilon [\nabla_{\rho}\nabla_{\r_1}\dots\nabla_{\r_m}R_{\sigma\delta\epsilon\nu}]\, \,  S_{\alpha\beta}{}^{\tau\l\r_1\dots\r_m\g\delta\m\n}|_\S \, .
 \ee

The same computation as in the previous section proves that $\tau_{\a\b}$ is tangent to $\S$ 
$$
n^\a \tau_{\a\b} =0 .
$$
Similarly, the generalized Israel conditions \eqref{genIsrael}  hold again, and the comments after \eqref{genIsrael} are still valid.

The case of a {\em proper matching} with only finite jumps of the energy-momentum tensor require the extra condition $\tau_{\a\b}=0$ so that generically, in addition to \eqref{manydisc=0}, one needs one extra continuity condition
\be\label{Manydisc=0}
[\nabla_{\a_1}\dots \nabla_{\a_i} R_{\a\b\m\n}] =0 \hspace{1cm} \mbox{for all} \hspace{1cm} i\in\{0,1,\dots,m+1\} .
\ee
together with \eqref{Kdisc=0}. The same warning as before is needed, as the vanishing of $\tau_{\a\b}$ may only require part of $[\nabla_{\a_1}\dots \nabla_{\a_{m+1}} R_{\a\b\m\n}] $ to vanish for very specific expressions of $\hat F$.

\section{Discussion}
The main results in this paper are the general formulas for $\tau_{\mu\nu}$ Eqs.\eqref{tau}-\eqref{tau2}, the generalized Israel relations \eqref{genIsrael} and the matching conditions \eqref{nT}, and the proof that \eqref{genIsrael} and \eqref{nT} are valid in completely general (diffeomorphism invariant) theories. 

The results have been derived using the theory of tensor distributions. More general conditions could be sought by resorting to the theory of generalized functions \cite{SV,GKOS}, where some non-linear products of distributions can be computed within some algebras. This has actually been explored in \cite{CT}. However, as per today this approach is not free of indeterminacies, so that I have preferred to stick to the linear distributional theory to be on the safe side. 

The minimal coupling of matter has been assumed throughout. More general couplings can be analyzed using the same techniques and guiding principles, as has been done in scalar-tensor theories for example \cite{AMM}. However, the impossibility of treating the couplings in general makes it impossible to derive similar generic expressions in that situation. As another remark, our analysis relies on the metric theory, not considering the connection as independent variable. For the case of Palatini type theories, the same distributional approach can be used, see e.g. \cite{OR}.

When the Lagrangian density depends on derivatives of the curvature, I have assumed that there is a maximum order derivative arising in the action. There are cases with infinite derivative Lagrangians though, and they are more difficult to study. In principle, the Riemann tensor field should be infinitely differentiable, but there may be situations in which other conditions can be proposed. An example was explored in \cite{KMM}, where the authors needed to add ad hoc assumptions to make sense of the junction. 

As a final comment, I would like to stress that if the matter contents of the problem under consideration is identified, then the corresponding field equations and junction conditions for the fields involved must be added to the picture, as otherwise undesired indeterminacies arise \cite{FMS}.

\section*{Acknowledgments}
Work done during a visit to the Mathematics Department of Nagoya University.
Support from the Grant PID2021-123226NB-I00 funded by the Spanish MCIN/AEI/10.13039/501100011033
together with “ERDF A way of making Europe” and by Spanish MICINN Project No. PID2021-126217NB-I00. 

\appendix

\section{Appendix}
Once the preliminary junction conditions \eqref{h=h} have been implemented we have, as explained in the main text, a unique normal $n_\mu$ and a basis of vector fields $\{\vec e_a\}$ on $\Sigma$ such that \eqref{h=h} holds and a continuous metric $g_{\m\n}$ exists on the spacetime.
The space-time version of the first fundamental form, unique due to (\ref{h=h}), is the projector to $\Sigma$ (defined only on $\S$)
\be
h_{\mu\nu}=g_{\mu\nu}-n_\mu n_\nu \label{proj}
\ee
so that
$$
n^\mu h_{\mu\nu} =0, \hspace{1cm} h_{\mu\rho} h^\rho{}_\nu =h_{\mu\nu}, \hspace{1cm} h^\mu{}_\mu =n, \hspace{1cm}  h_{\mu\nu}e^\mu_a e^\nu_b =h_{ab}
$$
$$
e^\mu_a =h_{ab}\, \omega^b_\nu \, g^{\nu\mu} , \hspace{1cm} e^\mu_c \omega^c_\nu =h^\mu_\nu \, .
$$

The extrinsic curvatures, or second fundamental forms, inherited by $\S$ from both sides $M^\pm$ are given by
\be
K^\pm_{ab} \equiv -n_\mu e^\rho_a\nabla^\pm_\rho e^\mu_b =e^\mu_b  e^\rho_a\nabla^\pm_\rho n_\mu =-n^\pm_\mu \left(\frac{\partial^2 x^\mu_\pm}{\partial\xi^a\partial\xi^b}+\Gamma^{\pm \mu}_{\rho\sigma}\frac{\partial x_\pm^\rho}{\partial \xi^a} \frac{\partial x_\pm^\sigma}{\partial \xi^b}\right), \label{2FF}
\ee
and at this stage they may disagree on $\S$ because the derivatives of the metric are not continuous in general.
Their spacetime versions are denoted by $K^\pm_{\mu\nu}$ and defined by 
$$
K^\pm_{\mu\nu}\defi \omega^a_\mu \omega^b_\nu K^\pm_{ab} =h^\rho{}_{\nu}h^\sigma_\mu \nabla^\pm_\rho n_\sigma \hspace{1cm} K^\pm_{\mu\nu}=K^\pm_{\nu\mu} .
$$
Trivially $n^\mu K^\pm_{\mu\nu}=0$, thus only the $n(n+1)/2$ components  tangent to $\Sigma$ are non-identically vanishing.

In order to compute the curvature tensor field, and given that the metric is only continuous, one resorts to using the theory of tensor distributions \cite{L,SV,MS,RSV}. To that end one introduces the $\Sigma$-step function $\theta : M \rightarrow  \mathbb{R} $ as
\bea
\theta =\left\{
\begin{array}{ccc}
1 &  & M^+\\
1/2 & \mbox{on} & \Sigma \\
0 &  & M^-
\end{array}\right. \label{theta}
\eea
This is locally integrable so that it defines a scalar distribution $\otheta$,\footnote{I distinguish between the object seen as a tensor (or scalar) field and the object seen as a tensor (or scalar) distribution by placing an underline on the distribution version.}
whose covariant derivative is a one-form distribution with support on $\S$ 
\be
\nabla_\mu\,   \underline{\theta} = \underline{\delta}_\mu =n_\mu\,  \deltaS \label{nablatheta}
\ee
where $\deltaS$ is a scalar distribution acting on test functions $Y$ by
\be
\left < \deltaS, Y \right >\defi \int_{\Sigma} Y \enspace .\label{deltaS}
\ee

Consider any tensor field $T$ which (i) may be discontinuous across $\Sigma$, (ii) is differentiable
on $M^{+}$ and $M^{-}$, and (iii) $T$ and its covariant derivative have
definite limits on $\Sigma$ coming from both sides $M^\pm$. The restrictions of $T$ to $M^{\pm}$ are denoted by $T^{\pm}$, respectively. The tensor distribution associated to such a $T$ exists and is denoted by
\be
\underline{T}=T^{+} \otheta +T^{-} \left( \1-\otheta \right ) \enspace . \label{Tdist}
\ee
By definition, we also take
$$
T\defi T^{+} \theta +T^{-} \left( 1-\theta \right )
$$
as the tensor field defined {\em everywhere}, and thus at each point of $\S$
\be
T^\S\defi T|_\S = \frac{1}{2} \left(\mathop{\lim} \limits_{x \mathop \to \limits_{M^{+}}\S} T^{+}(x) +
\mathop{\lim} \limits_{x \mathop \to \limits_{M^{-}}\S} T^{-}(x) \right)\, .\label{TonS}
\ee
The covariant derivative of $\underline T$ is \cite{MS,RSV}
\be
\nabla_\mu\, \underline{T}^{\a_1\dots\a_q}_{\b_1\dots\b_p} = \nabla_\mu T^{+\a_1\dots\a_q}_{\b_1\dots\b_p} \otheta
+\nabla_\mu T^{-\a_1\dots\a_q}_{\b_1\dots\b_p} (\1 -\otheta )+\left[T^{\a_1\dots\a_q}_{\b_1\dots\b_p} \right] n_\mu \deltaS
\label{nablaT1}
\ee
where
\be
\forall q \in \Sigma, \hspace{1cm} \ \left[T^{\a_1\dots\a_q}_{\b_1\dots\b_p} \right]  \equiv
 \mathop{\lim} \limits_{x \mathop \to \limits_{M^{+}}q} T^{+}{}^{\a_1\dots\a_q}_{\b_1\dots\b_p}(x) -
\mathop{\lim} \limits_{x \mathop \to \limits_{M^{-}}q} T^{-}{}^{\a_1\dots\a_q}_{\b_1\dots\b_p}(x) \enspace 
\label{discont}
\ee
are the components of a tensor field defined only on $\Sigma$, that is called the ``jump'' or the ``discontinuity'' of $T$ at $\Sigma$. Formula (\ref{nablatheta}) is just a particular case of this general formula.

In particular, the Riemann tensor distribution can be computed and the result is
\be
\underline{R}^\alpha{}_{\beta\mu\nu}=R^{+\alpha}{}_{\beta\mu\nu}\otheta + R^{-\alpha}{}_{\beta\mu\nu}(\1-\underline{\theta})+H^\alpha{}_{\beta\mu\nu}\, \deltaS \label{Riedist}
\ee
where 
\be
H_{\alpha\beta\lambda\mu} = - n_{\alpha}\left[K_{\beta\mu}\right]n_{\lambda}+n_{\alpha}\left[K_{\beta\lambda}\right]n_{\mu}-n_{\beta}\left[K_{\alpha\lambda}\right]n_{\mu}+n_{\beta}\left[K_{\alpha\mu}\right]n_{\lambda} \label{HRie}
\ee
is called the singular part of the Riemann tensor distribution and depends exclusively on the {\em jump} on $\Sigma$ of the second fundamental form 
\be
      \left[K_{\mu\nu}\right]\defi K^{+}_{\mu\nu}-K^{-}_{\mu\nu}, \hspace{1cm} n^{\mu}\left[K_{\mu\nu}\right] =0 \label{Kdisc}.
\ee
From here one also obtains the Ricci tensor distribution
\be
\underline{R}_{\b\m}=R^+_{\b\m}\otheta +R^-_{\b\m} (\1-\otheta) +H_{\b\m} \deltaS \label{Ricdist}
\ee
\be
H_{\beta\mu}\defi H^\rho{}_{\beta\rho\mu} =-\left[K_{\beta\mu}\right] -\left[K^\rho{}_{\rho}\right] n_\beta n_\mu ;\label{HRic}
\ee
the scalar curvature distribution
\be
\underline R = R^+\otheta +R^- (\1-\otheta) + H \deltaS \label{scalardist}
\ee
\be
H\defi H^\rho{}_{\rho}=-2\left[K^\mu{}_{\mu}\right]  ; \label{Hscalar}
\ee
and the Einstein tensor distribution
\be
\underline{G}_{\b\m} \defi \underline{R}_{\b\m}-\frac{1}{2} g_{\b\m} \underline R =G^+_{\b\m}\otheta +G^-_{\b\m} (\1-\otheta) +{\cal G}_{\b\m} \deltaS \label{Gdist}
\ee
\be
{\cal G}_{\beta\mu} = -\left[K_{\beta\mu}\right]+h_{\beta\mu}\left[K^\rho{}_{\rho}\right] , \hspace{1cm} n^\mu {\cal G}_{\beta\mu} =0 \label{HG} .
\ee
Observe that the singular part of the Ricci tensor distribution vanishes if, and only if, the jump of the second fundamental form vanishes, hence, if and only if that of the full Riemann tensor distribution does. 

The first covariant derivative of $\underline{R}^\alpha{}_{\beta\mu\nu}$ reads
\be\label{nablaRie}
\nabla_\l \underline{R}^\alpha{}_{\beta\mu\nu}=\nabla_\l R^{+\a}{}_{\b\m\n} \otheta +\nabla_\l R^{-\a}{}_{\b\m\n} (\1-\otheta) + \left[R^\a{}_{\b\m\n}\right] n_\l \deltaS +\nabla_\l (H^\alpha{}_{\beta\mu\nu}\, \deltaS)
\ee
and the second Bianchi identity holds in the distributional sense:
$$
\nabla_{\rho}\underline{R}^\alpha{}_{\beta\mu\nu}+\nabla_{\mu}\underline{R}^\alpha{}_{\beta\nu\rho}+\nabla_{\nu}\underline{R}^\alpha{}_{\beta\rho\mu}=0
$$ 
from where one deduces by contraction
$$
\nabla^\beta \underline{G}_{\beta\mu}=0
$$
for the Einstein tensor distribution. By using (\ref{Gdist}) and the general formula (\ref{nablaT1}) this implies \cite{RSV,S4}
\be
0= \nabla^\beta \underline{G}_{\beta\mu} = n^\beta \left[G_{\beta\mu}\right] \underline{\delta}^\Sigma +\nabla^\beta \left({\cal G}_{\beta\mu}\underline{\delta}^\Sigma \right)
=\underline{\delta}^\Sigma\left(n^\beta \left[G_{\beta\mu}\right] +\overline\nabla^\beta {\cal G}_{\beta\mu}-\frac{1}{2}n_\mu {\cal G}^{\rho\sigma}(K^+_{\rho\sigma}+K^-_{\rho\sigma})\right)\, \label{divG=0}
\ee
or equivalently, by splitting into normal and tangent components,
\bea
(K^+_{\rho\sigma}+K^-_{\rho\sigma}){\cal G}^{\rho\sigma} = 2n^\beta n^\mu \left[ G_{\beta\mu}\right]=2n^\beta n^\mu \left[ R_{\beta\mu}\right]-[R], \label{1}\\
\overline\nabla^\beta {\cal G}_{\beta\mu}=-n^\rho h^\sigma{}_\mu \left[ G_{\rho\sigma}\right]=-n^\rho h^\sigma{}_\mu \left[ R_{\rho\sigma}\right] \label{2} .
\eea

In the important case (and the main one analyzed in this paper) in which \eqref{Kdisc=0} holds,
\eqref{nablaRie} reduces to 
\be\label{nablaRie1}
\nabla_\l \underline{R}^\alpha{}_{\beta\mu\nu}=\nabla_\l R^{+\a}{}_{\b\m\n} \otheta +\nabla_\l R^{-\a}{}_{\b\m\n} (\1-\otheta) + \left[R^\a{}_{\b\m\n}\right] n_\l \deltaS \, .
\ee
Under the same assumption \eqref{Kdisc=0} a standard calculation \cite{MS,RSV}  leads to
\begin{equation}
  \label{eq:[Rie]}
  [R_{\alpha\beta\lambda\mu}]=n_\alpha n_\lambda B_{\beta\mu}-n_\lambda n_\beta B_{\alpha\mu}-n_\mu n_\alpha B_{\beta\lambda}+n_\mu n_\beta B_{\alpha\lambda}
\end{equation}
where $B_{\alpha\beta}$ is an arbitrary symmetric tensor field tangent to $\Sigma$,
$$
B_{\alpha\beta}n^\alpha=0 \hspace{1cm} \Longrightarrow \hspace{3mm} B_{\a\b} = B_{ab}\omega^a_\a \omega^b_\b 
$$
that contains the $(n^2+n)/2$ allowed independent discontinuities for the curvature tensor.
Elementary contractions on (\ref{eq:[Rie]}) give
\begin{equation}
  \label{eq:[ricci]}
  [R_{\beta\mu}]=B_{\beta\mu}+\frac{1}{2} n_\beta n_\mu [R],\qquad [R]=2B^\r_\r,
\end{equation}
or equivalently
\be
B_{\b\m} =   [R_{\beta\mu}]- \frac{1}{2}[R] n_\beta n_\mu =[G_{\b\m}] +\frac{1}{2} h_{\b\m} [R] \, .\label{B=G}
\ee
In other words, the $n(n+1)/2$ allowed independent discontinuities of the Riemann tensor can be chosen to be the discontinuities of the $\S$-tangent part of the Einstein tensor (or equivalently, of the Ricci tensor).

In order to take further covariant derivatives of the above tensor distributions, one needs to take derivatives of tensor distributions of type
$$
t_{\a_1\dots \a_p} \deltaS
$$
where $t_{\a_1\dots \a_p}$ is any tensor field defined {\em at least} on $\S$, but not necessarily off $\S$. How to deal with these derivatives was shown in \cite{RSV}  and the general formula reads
\be
\nabla_\l \left(t_{\a_1\dots \a_p}  \deltaS \right) =\nabla_\r\left(t_{\a_1\dots \a_p}  n_\l n^\r \deltaS\right)+
 \left( h^\r_\l \nabla_\r t_{\a_1\dots \a_p} -K^{\S\r}{}_\r \, n_\l t_{\a_1\dots \a_p}\right)\deltaS \label{nablaTdelta} .
\ee

With the assumption \eqref{Kdisc=0} in place, a further covariant derivative of the Riemann curvature can be computed, and applying \eqref{nablaT1} to \eqref{nablaRie1} one gets
\be\label{nablanablaRie}
\nabla_\tau \nabla_\l \underline{R}^\alpha{}_{\beta\mu\nu}=\nabla_\tau\nabla_\l R^{+\a}{}_{\b\m\n} \otheta +\nabla_\tau\nabla_\l R^{-\a}{}_{\b\m\n} (\1-\otheta) + \left[\nabla_\l R^\a{}_{\b\m\n}\right] n_\tau \deltaS + \nabla_\rho(\left[R^\a{}_{\b\m\n}\right] n_\l n_\tau n^\rho \deltaS )
\ee
where use of the general \eqref{nablaTdelta} has been made in the last summand. This involves the discontinuities of the covariant derivative of the curvature tensor that can be computed as
\begin{equation}
  \label{eq:[d_Rie]}
  [\nabla_{\rho}R_{\alpha\beta\lambda\mu}]=n_\rho r_{\alpha\beta\lambda\mu}+ h^{\sigma}_\rho\nabla_\sigma[R_{\alpha\beta\lambda\mu}],
\end{equation}
where $r_{\alpha\beta\mu\nu}$ is the following tensor field (defined only on $\S$) with the symmetries of a Riemann
tensor \cite{MS,RSV}: 
\begin{eqnarray}
r_{\alpha\beta\m\n}&=& K_{\alpha\mu}B_{\nu\beta}
-K_{\alpha\nu}B_{\mu\beta}+K_{\beta\nu}B_{\mu\alpha}-K_{\beta\mu}B_{\nu\alpha}\nonumber\\
&+&\left(\Nb_\mu B_{\rho\nu}-\Nb_\nu B_{\rho\mu}\right)(n_\alpha h^\rho_\beta-n_\beta h^\rho_\alpha)
+\left(\Nb_\a B_{\rho\b}-\Nb_\b B_{\rho\a}\right)(n_\m h^\rho_\n-n_\n h^\rho_\m)\nonumber\\
&+& n_\a n_\m \r_{\b\n} -n_\a n_\n \r_{\b\m} - n_\b n_\m \r_{\a\n}+ n_\b n_\n \r_{\a\m}, \label{eq:rel_tB}
\end{eqnarray}
where $\r_{\b\m}$ is a new symmetric tensor field defined only on $\S$,  tangent to $\S$ ($n^\b\r_{\b\m}=0$), encoding the new $n(n+1)/2$ independent discontinuities of 
$\nabla_{\rho}R_{\alpha\beta\lambda\mu}$.

One can proceed in the same manner if further covariant derivatives of the Riemann tensor distribution are required. However, one must keep in mind that products of such derivatives with themselves will generically be ill-defined (in the sense of distributions) unless they do not possess any singular part (i.e., any part proportional to $\deltaS$ or to derivatives involving $\deltaS$): in other words, unless they are associated to a locally integrable tensor field. This is particularly relevant for the case of gravitational theories with actions that depend on invariants constructed using derivatives of the curvature.

{\bf Remark:}  An important remark is that the formulae in this appendix are {\em purely geometric}, independent of any field equations, and therefore valid in any theory of gravity based on a Lorentzian manifold.

\end{document}